# Theoretical modeling of synergistic effect of pores and grains on transmittance in transparent piezoelectric ceramics

Zixiang Xiong[1,2], Jian Zhu[1,2], Xueqian Geng[1], Zheng Wen[3], Yongcheng Zhang[1,2*], and Jianyi Liu[1,2†]

[1]*Centre for Theoretical and Computational Physics, College of Physics, Qingdao University, Qingdao 266071, China*

[2]*College of Physics, University-Industry Joint Center for Ocean Observation and Broadband Communication, National Demonstration Center for Experimental Applied Physics Education, Qingdao University, Qingdao 266071, China*

[3]*College of Electronics and Information, Qingdao University, Qingdao 266071, China.*

*Corresponding author: qdzhyc@qdu.edu.cn and jianyi_liu@qdu.edu.cn

ABSTRACT

Transparent piezoelectric ceramics (TPCs) have great application potential in electro-optical-mechanical multi-functional devices. Preparing high-performance TPCs, especially improving the transparency through microstructure regulation, has recently caused extensive discussion. However, there is still controversy about the influence of grains and pores on the transmittance of ceramics, and there is arbitrariness in the estimation of the theoretical transmittance limit. In this paper, taking PMN-PT-based ceramics as an example, theoretical mechanisms for the transmittance are discussed. An inhomogeneous reflection model is derived to improve the theoretical limit of transmittance. The effects of pores and grains scattering on transmittance are investigated. Rayleigh and RGD approximation are discussed to reveal the underlying scattering mechanisms. It is found that Rayleigh approximation is suitable for describing pore scattering, while RGD approximation is suitable for grain scattering. Thus, a Rayleigh-RGD combined model is proposed to describe light scattering in TPCs and successfully employed to fit experimentally measured transmittance curves.

# I. INTRODUCTION

Transparent piezoelectric materials have long been of research interest in the past decades due to their combination of optical transparency with piezoelectricity, which is quite promising for developing electro-optical-mechanical multi-functional devices, e.g., photoacoustic imaging sensors [1], transparent ultrasonic transducers [2] and electro-optical modulators [3], etc. However, it is generally considered difficult to obtain both perfect transparency and huge piezoelectricity in one material since most high-performance piezoelectric contain high density light scattering microstructures such as domain walls. Researchers typically have to compromise between piezoelectricity and transparency in material preparation [4-6]. Recently, Li and co-workers [7] reported using alternating-current electric field to regulate the microstructures of $Pb(Mg_{1/3}Nb_{2/3})O_3$-$xPbTiO_3$ (PMN-PT) single crystals to generate high performance transparent piezoelectric crystals. Their work suggested an opportunity for breaking the restrictive relationship between transparency and piezoelectricity and quickly sparked lots of investigations to prepare more high performance transparent piezoelectric ceramics. Yet polycrystalline ceramics possess more diverse microstructures, which results in more complex underlying mechanisms for transparency. It remains challenging to obtain excellent optical transparency in piezoelectric ceramics.

A commonly used strategy nowadays is to ensure better transparency in traditional piezoelectric ceramics by refining the grain size and reducing the porosity [8-11]. It is generally considered that anisotropic piezoelectric ceramics with smaller grain sizes and lower porosity do achieve higher transparency by reducing light scattering. This, however, is not always the case. There are publications [12-14] reported the preparation of PMN-PT based ceramics with various grain size ranging from about 4 μm to above 10 μm, yet their experimentally measured transmittances are all close to the theoretical limit. Moreover, low porosity is a necessary condition for transparent ceramics, and it is widely considered that when the total porosity exceeds 1%, a polycrystalline ceramic is completely opaque no matter how favorable other conditions are [15]. On the other hand, there are also investigations [16] prepared ceramics with relative densities less than 99% and quite good transmittance. All these results suggest that other mechanisms are at work underneath.

Theoretically, there are ongoing attempts to explain the experimental findings and reveal the underlying mechanisms governing transparency in ceramics [10,17,18], most of which are focused on traditional transparent ceramics like $Al_2O_3$ [19], $ZrO_2$ [20], $MgF_2$ [21], $Y_2O_3$ [22], and $MgAl_2O_4$ [23], etc. Theoretical studies on the transparency of piezoelectric ceramics such as PMN-PT with more complex microstructures are still lacking. As a result, the predictions directly based on the theories for above transparent ceramics often do not match the experimental results of piezoelectric ceramics. For example, it is surprising to find that the experimentally measured transmittance of piezoelectric ceramics is often equal to or even higher than the predicted limit based on traditional transparency theory [24-26]. Thus, a dedicated theoretical model for the transmittance of transparent piezoelectric ceramics is needed. The aim of this paper is to take PMN-PT ceramics as an example, reconsider reflection, scattering, absorption and other factors, and establish a transparency theory for transparent piezoelectric ceramics. Based on the renewed theory, the influence of pores and grains on the transmittance of transparent piezoelectric ceramics is revealed to provide theoretical support for the experimental preparation of high-performance transparent piezoelectric ceramics.

## II. THEORETICAL MODEL

As shown in Fig. 1, in order to quantitatively describe the transparency of ceramics, we define a physical quantity called transmittance. When measuring transmittance experimentally, the incident light is perpendicular to the sample surface, and the light detectors in spectrophotometer usually capture light under angles of up to 3-5°. Such measured transmittance is the so-called in-line transmittance. It is clear that a certain amount of the scattered light is captured by the detector in such cases, that give rise to seemingly high levels of transmittance. To give valid parameter for transparency quantification, additional treatment is usually adopted to make sure the transmittance measured under 0.5°, which is known as the real in-line transmittance (RIT). It is notable that here in our paper the theoretical maximum transmittance $T_{max}$ is calculated strictly in the direction of the incident light, which is an ideal situation.

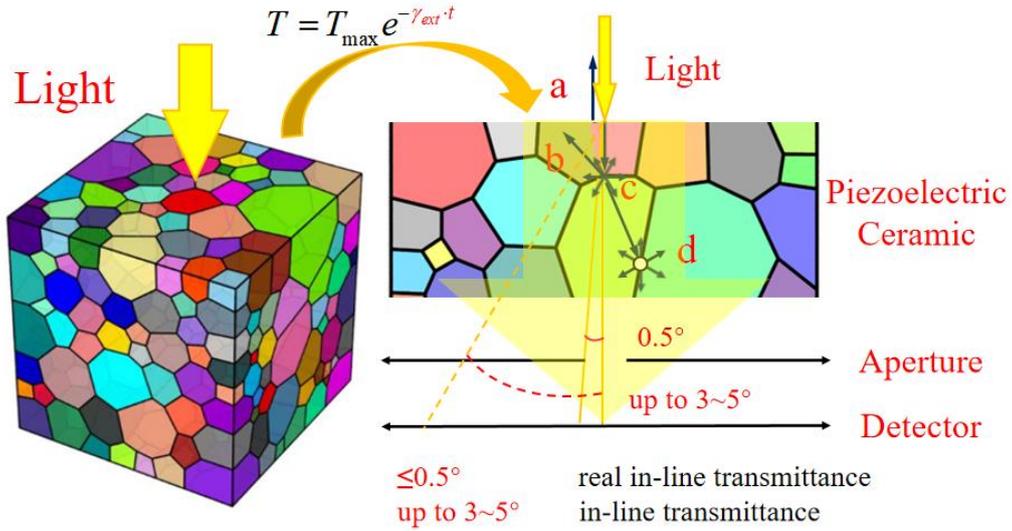

Fig. 1 Schematic diagram of the transmittance measurement model.

The transmittance of ceramics is governed by a combination of intrinsic and extrinsic factors. The intrinsic factors include specular reflection and light absorption, and the extrinsic factors are refraction, light scattering by pores, grains, and impurities. By ignoring the refraction [15], we start with the generalized model for determining the theoretical maximum transmittance $T$ of transparent ceramics [27]

$$T = (1-R_s)\exp\left[-(\gamma+\alpha)*t\right] \quad (1)$$

Here $R_s$ is the total reflection loss, $\gamma$ is the total scattering coefficient, $\alpha$ is the absorption coefficient, and $t$ is the thickness of the ceramic sample.

*A. REFLECTION*

The reflection loss $R_s$ is usually evaluated based on simplified models where the transparent ceramic sample is treated as homogeneous thus the incident light first enters the sample through the air-sample top interface then passes through the homogeneous sample and exits through the sample-air bottom interface. If we further ignore the multiple reflections between the top and bottom interfaces, $R_s$ then can be easily calculated according to Fresnel's formula

$$R_s = 1 - \left[\frac{4n_1 n_2}{(n_1 + n_2)^2}\right]^2 \tag{2}$$

where $n_1$ ($n_1 = 1$) and $n_2$ are the refractive index of air and the transparent ceramic sample. For PMN-PT ceramics, $n_2$ is reported to be between 2.3 and 2.6 [28]. Taking $n_2 = 2.3$, we can get $R_s = 0.286$ and the theoretical maximum transmittance $T_{max} = 1 - R_s = 0.714$ following Eq. (2). We note that this estimated $R_s$ based on the Single Reflection Model (SRM) is the theoretical limit value widely used in many available PMN-PT experimental publications when calculating reflection losses [29]. However, the single reflection model used to derive Eq. (2) is oversimplified. First, the multiple reflections between the top and bottom interfaces do reduce the reflection loss, thus cannot be neglected. Second, the ceramic sample is in fact inhomogeneous due to the anisotropy of grains[21,30] and the reflections between the grain interfaces should be discussed in the reflection model.

By taking into account the multiple reflections to develop a Multiple Reflection Model (MRM), we can get the modified $R_s$ as

$$R_s = \frac{2R'}{1+R'} \quad \text{with} \quad R' = \left(\frac{n-1}{n+1}\right)^2 \tag{3}$$

Likewise, taking $n_2 = 2.3$, we have $R_s = 0.269$ and the theoretical maximum transmittance $T_{max} = 0.731$. As a result, the evaluated theoretical limit increases by 2.4% when considering multiple reflections between the top and bottom interfaces. Furthermore, we consider the sample as inhomogeneous and develop an

Inhomogeneous Reflection Model (IRM). If multiple reflections between grains are also taken into account, and the renewed $R_s$ is written as (see the Supplementary material for detailed derivation process)

$$R_s = \prod[R_l^1 \prod(R_l^2)] \qquad (4)$$

where $R_l^1 = \dfrac{t_l r_{l+1}}{1 - r_l r_{l+1}}$ and $R_l^2 = \dfrac{1 - r_l}{1 - r_l r_{l+1}}$ represent the reflected light reflected to the previous grain and the reflected light reflected again to the previous grain. $t_l$ and $r_l$ represent the transmitted and reflected light differentiated by light passing through the $l^{th}$ grain interface. According to Eq. (4), assume that there are 100 grains along the direction of light incidence. Since the orientations of the grains are different, the refractive index of the grains is different in the direction of light incidence. Accordingly, we set the average refractive index of all grains in the sample being 2.3 and the maximum value of birefringence is 0.02, thus the refractive index of each grain is between 2.28 and 2.32 along the incident light direction. Note that the final reflection loss by Eq. (4) is not a fixed value but is closely related to the distribution of the refractive index (orientation of the grains).

In Fig. 2, we plot the theoretical maximum transmittance predicted by single reflection model, multiple reflection model and inhomogeneous reflection model at different incident light wavelengths. The effect of light absorption is also considered here (see ABSORPTION section). Fig. 2 shows multiple reflections do significantly increase the theoretical transmittance limit, while the distribution of grain orientation will slightly increase or decrease the transmittance limit.

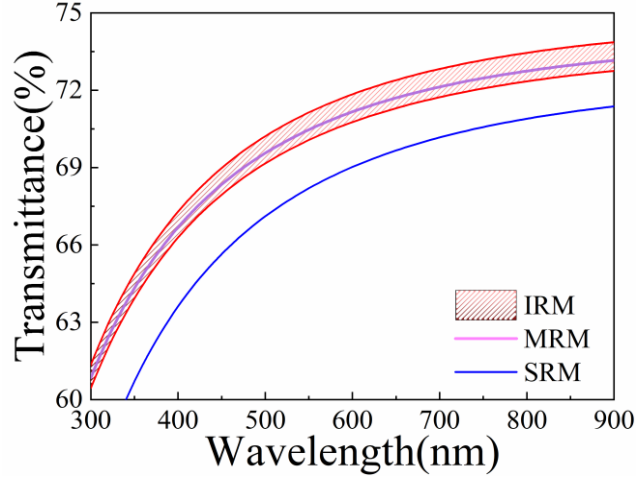

Fig. 2 Comparison of transmittance theory limits of Single Reflection Model (SRM), Multiple Reflection Model (MRM) and Inhomogeneous Reflection Model (IRM).

B. SCATTERING

Reflection determines the theoretical maximum transmittance, while scattering governs the actual transmittance. As shown in Eq. (1), the effect of scattering on the transmittance can be described by the Lambert-Beer extinction law [31]

$$T = T_{\max} \exp(-\gamma t) \quad (5)$$

and the scattering coefficient $\gamma$ is written as $\gamma = N_V C_{sca}$, where $N_V$ is the number of scatterers (grains or pores) per unit volume, and $C_{sca}$ is the scattering cross section. To be more simplified, assume the shape of the scatters are spherical and we can get

$$\gamma_{sca} = \frac{3\phi}{2d} Q_{sca} \quad (6)$$

Here $\phi$ and $d$ are the volume fraction and diameter of the scatters, respectively. $Q_{sca}$ is the scattering efficiency.

Mie in 1908 [32] developed an exact solution to the problem of electromagnetic wave scattering by a spherical, birefringent, isotropic scatterer (possibly an absorber) in a

non-absorbing medium, where $Q_{sca}$ is written as

$$Q_{sca}^{Mie} = \frac{2}{x^2} \sum_{j=1}^{\infty} (2j+1) \text{Re}(a_j + b_j) \tag{7}$$

where $x$ is the relative size parameter, i.e., $x = \pi d n_{med}/\lambda_0$ with $n_{med}$ being the refractive index of the medium and $\lambda_0$ being the wavelength of light in vacuum. The coefficients $a_j$ and $b_j$ are functions of the relative size and the relative (complex) index of refraction of the scatterers. It is notable that the numerical solution of Mie's formula depends on computer calculations. Although such calculations are not difficult nowadays, Mie theory is not perfect because physics, especially the relationship between $Q_{sca}$ and scattering factors like grain size $d$ and birefringence difference $\Delta n$ in Mie's formula (see Eq. (7)) is not intuitive. Based on Mie's theory, several approximate and simplified models are further developed [23]. For example, in the Rayleigh approximation, $Q_{sca}$ is approximated as

$$Q_{sca}^{Rayleigh} = \frac{8}{3}\left(\frac{m^2-1}{m^2+2}\right)^2 x^4 \tag{8}$$

with $m = n_{sca}/n_{med}$, where $n_{sca}$ is the refractive index of the scatterer, thus $\gamma$ is written as

$$\gamma^{Rayleigh} = \frac{4\pi^2 n_0^4}{\lambda_0^4} \phi d_{sca}^3 \frac{m^2-1}{m^2+2} \tag{9}$$

which is proportional to $\phi$, $d^3$, while its relationship to $\Delta n$ is implicit. In the Rayleigh-Gans-Debye (RGD) approximation, $Q_{sca}$ is approximated as

$$Q_{sca}^{RGD} = 2(m-1)^2 x^2 \tag{10}$$

and $\gamma$ is written as

$$\gamma^{RGD} = \frac{3\pi^2 \phi d_{sca}(n_{sca}-n_{med})^2}{\lambda_0^2} \tag{11}$$

which is proportional to $\phi$, $d$ and $(\Delta n)^2$, with $\Delta n = (n_{sca} - n_{med})$. Later in the article, we will consider Mie's model and the above two approximations and discuss their scope of applicability.

As for the choice of the value of $\phi$, Pabst et al. classified the use of existing scattering models into two categories: equivalent composite models (ECM) and dense polycrystalline models (DPM) [23]. ECM treats scatterers with different refractive indices as different scatterers ($\phi = 1/3$), and DPM considers the refractive index as the overall average refractive index of the material ($\phi = 0.27778$). Since there is only the refractive index of the ceramic and not the individual grains, ECM applies to pore factor effects, while DPM applies to the grain scale. Finally, for the pore effect, we selected $\phi$ values of 1E-2 to 1E-5 (porosity) and $n_{sca} = 1$ (vacuum refractive index). For the grain effect, we chose a $\phi = 0.27778$ and $n_{sca}$ as the refractive index for different grains.

## C. ABSORPTION

Generally speaking, the absorption coefficient can be calculated by the formula $\alpha = 4\pi k/\lambda_0$. $k$ is the extinction coefficient, indicating how hard the material absorbs light, which is also known as the attenuation or absorption coefficient. However, due to the complexity of the solid solution material components, the coefficients $\alpha$ or $k$ are difficult to determine. In this study, we fitted the light extinction coefficient $k \sim 3.77$ for PMN-PT materials based on the report of Qiu et al [7]. It is worth noting that k is difficult to derive before wavelengths of 400-450 nm due to the presence of intrinsic absorption bands (mainly from the jumps of bound electrons). We fit the $\alpha$ coefficient using a model of the form $\alpha = A*exp(-(\lambda + \Delta\lambda)/B)$ ($A = 1.09573E15$, $B = 12.76344$).

**III. RESULTS**

Based on the inhomogeneous reflection model, we further investigate the effects of pores and grains on the transmittance of the PMN-PT transparent ceramics based on Mie's theory for scattering in the following. The applicability and reliability of the Rayleigh model and the RGD model when dealing with the pores and grains effects on transmittance are also discussed.

*A. PORE EFFECTS*

When discussing the pore effect, we exclude the scattering of other factors and mainly consider the influence of porosity and pore size combing with sample thickness. We set the porosity range from 1E-2 to 1E-5, and the size of pores range from a few nanometers to 2 μm, and the thickness of the sample range from 0.3 mm to 1 mm, which covers most practical cases of transparent piezoelectric ceramic samples [25]. In addition, the wavelength is fixed at 900nm. The calculated results based on Mie's theory are shown in Fig. 3. In Fig. 3 (a), we fix the pore size as 100 nm and investigate the dependence of transmittance on porosity and the thickness of the sample. It shows that the transmittance increases with the decrease of porosity and thickness. In particular, in the parameters range we considered, porosity plays a decisive role in determining the transmittance compared to sample thickness. For the sample with a pore size of 100 nm, when the porosity is greater than 1E-3, the sample transmittance is less than 20%, which is impossible to achieve transparency. When the porosity is less than 1E-4, the sample transmittance is higher than 60%, even close to the theoretical limit of 73.7% given by our developed reflection model. By fixing the porosity as 1E-3, the calculated dependence of transmittance on pore size and the thickness of the sample is plotted in Fig. 3 (b). Interestingly, it can be found that the transmittance is not monotonically changed with the pore size. In specific, for the sample with a fixed porosity of 1E-3, when the pore size is less than 400 nm, the transmittance decreases rapidly from theoretical limit to nearly zero as the pore size increases. When the pore size is greater than 400 nm, the transmittance gradually increases with the increase of the pore size, approaching 50%. The critical pore size corresponding to the lowest value of transmittance is about 400 nm. This is in line with the findings on the pore size effect, where a "critical" pore size of about half the wavelength of light leads to the highest level of light scattering, approximately between 400~500 nm in the pore size. In Fig. 3 (c), we fixed the thickness of the transparent ceramic sample as 0.5 mm and comprehensively considered the effects of porosity and pore size on transmittance. The following are the main conclusions: First, when the pore size is less than 400 nm, the transmittance decreases rapidly with the increase of pore size and porosity. When the pore size is further less than 200 nm and the porosity is less than 1E-2, the transmittance of the sample is very close to the theoretical limit predicted by our developed reflection model. Second, when the pore size is greater than 400 nm, the transmittance decreases with the increase of porosity while increases with the increase of pore size. When the pore size is greater than 1.3 μm and the porosity is less than 1E-4, the transmittance of the sample can also be very close to the theoretical limit. Therefore, in order to reduce the influence of pore scattering on the transmittance of transparent ceramics, when preparing samples, it is very important to make the pore size smaller than 200 nm, or the porosity lower than 1E-4 according to our calculations.

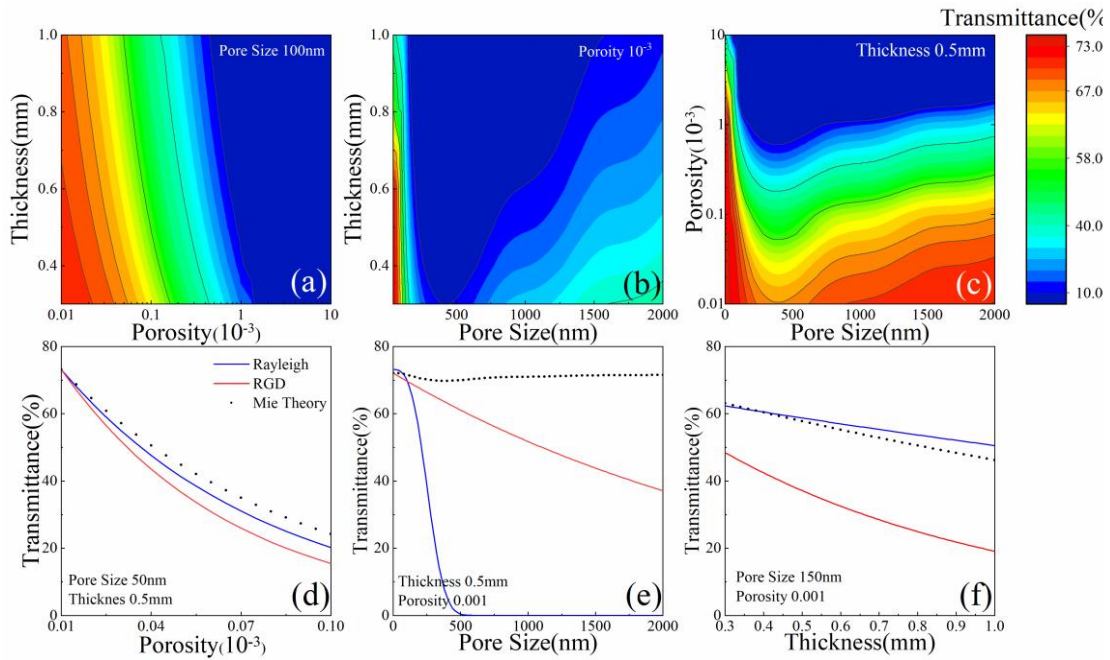

Fig. 3 (a)-(c) Calculation of the effect of different pore effects on transmittance based on Mie theory, including pore size, porosity and thickness. (e)-(f) Reliability and applicability of the Rayleigh approximation and the RGD approximation in the study of pore effects on the transmittance of transparent ceramics.

In addition, based on Mie's theory, we also studied the reliability and applicability of the commonly used Rayleigh approximation and RGD approximation in studying the influence of pore effect on the transmittance of transparent ceramics (see Fig. 3 (e)-(f)). Assuming the pore size as 200 nm and the sample thickness as 0.5 mm, both Rayleigh approximation and RGD approximation predict that the transmittance decreases with increasing porosity, which is consistent with the result by Mie's theory. However, the result of Rayleigh approximation is obviously closer to that of Mie theory, while the result of RGD approximation has a large deviation from Mie theory although the trend is the same. This conclusion is also correct when fixing the porosity as 1E-4 and pore size as 200 nm and studying the variance of the transmittance with sample thickness. It is worth noting that for a fixed porosity of 1E-4 and sample thickness of 0.5 mm, the situation is completely different when studying the variation of transmittance with pore size. When the pore size is greater than 200 nm, neither the Rayleigh approximation nor the RGD approximation can accurately describe the variation of transmittance with pore size, and even the variance trend cannot be correctly predicted. When the pore size is less than 200 nm, the Rayleigh approximation can give results consistent with the result by Mie's theory. In summary, it shows that Rayleigh approximation can better predict the influence of the pore effect on transmittance, and the scattering factor $\gamma$ is proportional to porosity $\phi$ and the cube of pore size, i.e., $d^3$ according to Eq. (9).

B. GRAIN EFFECTS

To discuss the effects of grain factors on transmittance of the transparent ceramics, we focus on the influence of birefringence difference $\Delta n$ and grain size combing with sample thickness. We set the birefringence difference $\Delta n$ range from 0 to 5E-3, and the size of pores range from 1 to 20 μm, and the thickness of the sample range from 0.3 mm to 1 mm, which covers most practical cases of transparent piezoelectric ceramic samples. Also, the wavelength is fixed at 900 nm. Fig. 4 shows the calculated results of grain effects based on Mie's theory. By fixing the grain size as 5 μm, we can see from Fig. 4 (a) that the transmittance increases with decreasing $\Delta n$ value and sample thickness. Specifically, when $\Delta n$ is less than 2E-3, the transmittance of the ceramic approaches the theoretical limit. When $\Delta n$ is greater than 3E-3, the transmittance decreases rapidly as $\Delta n$ increases. Fixing $\Delta n$, the transmittance decreases as increasing grain size (see Fig. 4 (b)). In Fig. 4 (c), we fixed the thickness of the transparent ceramic sample as 0.5 mm and comprehensively considered the effects of $\Delta n$ and grain size on transmittance. It shows that when $\Delta n$ is small, grain size has little effect on the transmittance of the sample considered, thus highly transparent PMN-PT transparent ceramics can be prepared regardless of large or small grain size. This explains why the experimentally prepared PMN-PT based ceramics with various grain sizes ranging from about 5 μm to above 10 μm can all achieve excellent transparency. Moreover, when $\Delta n$ increases to be larger than 1.2E-3, the effect of grain size on the transmittance gradually appears, and if $\Delta n$ further increases to be larger than 2E-3, only PMN-PT transparent ceramics with small grain sizes (< 5 μm) can achieve good transparency with a transmittance of above 65%.

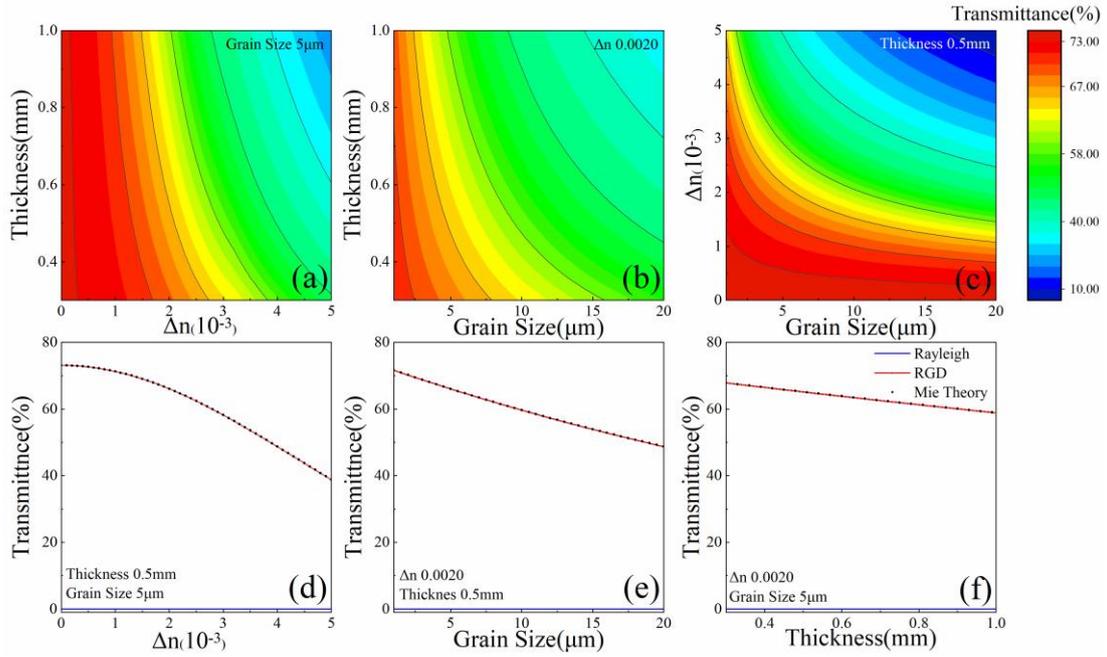

Fig. 4 Calculation of the effect of different grain effects on transmittance based on Mie theory, including grain size, $\Delta n$ and thickness. (e)-(f) Reliability and applicability of the Rayleigh approximation and the RGD approximation in the study of grain effects on the transmittance of transparent ceramics

Likewise, the reliability and applicability of the commonly used RGD approximation and Rayleigh approximation in studying the influence of grain effect on the transmittance of transparent ceramics are also discussed based on Mie's theory and the corresponding results are shown in Fig. 4 (e)-(f). We can see that the results of the RGD approximation are much closer to the results predicted by Mie's model, regardless of dealing with the effect of $\Delta n$, grain size or thickness on transmittance. In other words, the RGD approximation can better predict the effect of the grain effect on transmittance, and the scattering factor $\gamma$ is proportional to grain size $d$ and the square of birefringence difference $(\Delta n)^2$ according to Eq. (11).

*C. COMBINED EFFECTS*

In the following, we discuss the combined effect of pores and grains on the transmittance. The grain average grain size is set to be 5 μm, the birefringence difference $\Delta n$ is set to be 2E-3, the porosity is set as 1E-3, and the pore size is set as 25 nm. The thickness of the sample is 0.5 mm. In Fig. 5 (a), it shows the variation of transmittance with wavelength by Mie's theory. When only pore scattering is considered (black line), the transmittance increases with wavelength and reaches 0.72 at 900nm. When only grain scattering is considered (red line), the transmittance is lower than when only pores are considered in most bands and is 0.66 at 900nm. Obviously, for the sample considered, the loss caused by grain scattering is greater than pore scattering. When both scattering effects are considered (blue line), the transmittance further declines and is 0.64 at 900nm.

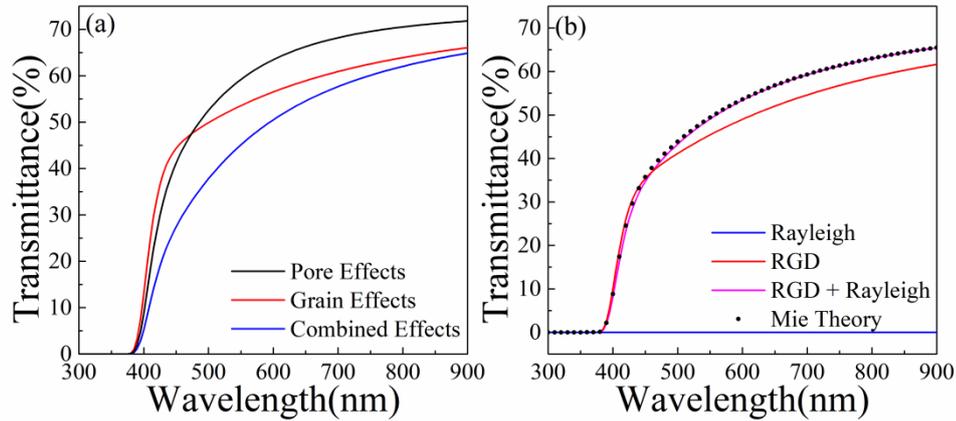

Fig. 5 (a) Comparison of transmittance under pore, grain, pore and grain scattering. (b) Compares the Rayleigh approximation, the RGD approximation, and the Rayleigh-RGD combination theory with the Mie theory

In Fig. 5 (b), Mie's theory, Rayleigh approximation, RGD approximation and Rayleigh-RGD combined theory are compared to predict the curves of PMN-PT-based ceramic transmittance changing with the wavelength of the incident light. Taking the results of Mie theory as the standard, the RGD approximation, according to our previous analysis, is more suitable for describing grain scattering and cannot accurately describe pore scattering. The predicted transmittance is significantly lower than the result of Mie's theory. The Rayleigh approximation, according to our previous analysis, is more suitable for describing pore scattering and cannot describe grain scattering. The predicted full-band transmittance is almost zero. This is because the Rayleigh approximation will produce a huge deviation when used to describe grain scattering as shown in Fig. 4 (e)-(f). Particularly, When Rayleigh-RGD combined approximation model

$$T = (1-R_s)T_\alpha T_{\text{Rayleigh}}^{\text{Pore}}(\phi d^3) T_{\text{RGD}}^{\text{Grain}}((\Delta n)^2 d) \tag{12}$$

where Rayleigh is used to describe pore scattering and RGD is used to describe grain scattering, is adopted, the predicted results are almost consistent with Mie's theory. Importantly, we can see that the mechanisms governing grain scattering and pore scattering are completely different. In addition, when the scattering of domain structures (without the scope of this work) is further taken into account, it can be imagined that there will be more abundant microscopic mechanisms. Therefore, it is important to point out that there are multiple possible mechanisms underneath governing the transparency of piezoelectric ceramics.

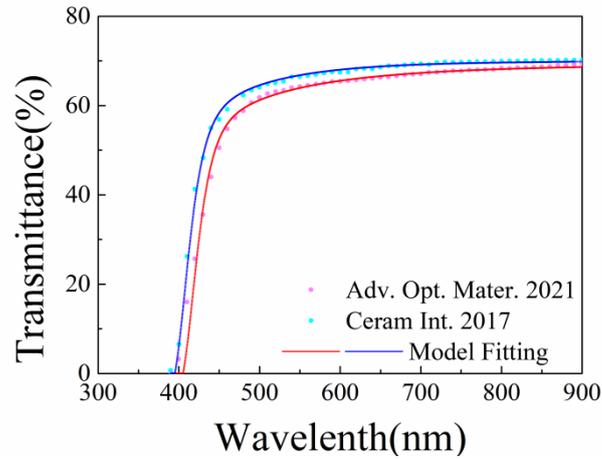

Fig. 6 the Rayleigh-RGD combined approximation model is adopted to fit the experimentally measured transmittance curve of PMN-PT-based transparent piezoelectric ceramics

In Fig. 6, the Rayleigh-RGD combined approximation model is adopted to fit the experimentally measured transmittance curve of PMN-PT-based transparent piezoelectric ceramics [24,25]. We can see that the theoretical and experimental results are in good agreement. In the Adv. Opt. Mater. experimental data points of Fig. 6 (blue dot), the obtained parameter of pores is $\phi d^3 = 4.94\text{E-}9$, and that of grains is $(\Delta n)^2 d = 1.59\text{E-}5$. Unfortunately, we usually do not have enough experimental data on pores and grains to compare with our theoretical predictions. Most of the time, we can only get the data on average grain size. Accordingly, the experimentally measured average grain size is 4.61 μm, thus it is easy to calculate the average value of birefringence difference $\Delta n$ being about 1.86E-3, which is within a reasonable range.

# IV CONCLUSION

In this paper, we studied the underneath mechanisms for the transmittance of transparent piezoelectric ceramics. An inhomogeneous reflection model was derived to improve the theoretical limit of transmittance. The effects of pore scattering and grain scattering on transmittance of transparent piezoelectric ceramics were investigated based on Mie's theory. Rayleigh approximation and RGD approximation were discussed to reveal the underlying scattering mechanisms. It is found that Rayleigh approximation is suitable for describing pore scattering, and the RGD approximation is suitable for describing grain scattering. Thus, a Rayleigh-RGD combined model was proposed to describe the light scattering in transparent piezoelectric ceramics. This model was then successfully employed to fit experimentally measured transmittance curve. Our study is promising to provide theoretical support for the experimental preparation of high-performance transparent piezoelectric ceramics.

# ACKNOWLEDGMENTS

This study was supported by the National Natural Science Foundation of China (NSFC) (Grants No. 12002400, No. 12132020, and No. 52272116) and the Taishan Scholar Program of Shandong Province (No. tstp20240511).